\documentclass[aps,pra,preprint,superscriptaddress,showpacs]{revtex4-1}

\usepackage{graphics}
\usepackage{float}
\usepackage{amsmath,amssymb}
\usepackage{longtable}
\usepackage{supertabular}
\usepackage[enableskew]{youngtab} 
\usepackage[pdftex]{graphicx}
\usepackage{color} 

\def\be{\begin{equation}}
\def\ee{\end{equation}}

\def\be{\begin{equation}}
\def\ee{\end{equation}}

\begin{document}
\DeclareGraphicsExtensions{.jpg,.pdf}

\title[On the application of Jucys-Murphy operators in the Hubbard model]{On the application of Jucys-Murphy operators in the Hubbard model}


\author{Dorota Jakubczyk}
\affiliation{Departament of Physics, Rzeszow University of Technology, al. Powsta\'nc\'ow Warszawy 12, 35-959 Rzesz\'ow, Poland}%
\author{Pawe\l\ Jakubczyk}
\affiliation{Faculty of Mathematics and Natural Sciences, University of Rzesz\'ow, Rejtana 16A, 35-310 Rzesz\'ow, Poland}%
\author{Yevgen Kravets}
\affiliation{Centre de Physique Th\'{e}orique (CPHT), \'{E}cole Polytechnique, 91128 Palaiseau, France}%


\date{\today}

\begin{abstract}
The operator techniques based on  the Jucys-Murphy operators {were} applied in the procedure of an immediate diagonalization of the one-dimensional Hubbard model. The Young orthogonal basis {was given by the} irreducible basis of the symmetric group acting on the set of nodes of the magnetic chain. The example of the attractive Hubbard model at the half-filled {magnetic rings} case was considered where the group $SU(2)\times SU(2)$ acts within the spin and pseudo-spin space. These techniques {significantly reduced} size of the eigenproblem of the Hubbard Hamiltonian.
\end{abstract}


\maketitle

\section{Introduction}

Hubbard model \cite{Hubbard}, is the simplest generalization beyond the band theory description of solids, allowing us to understand many interesting phenomena of the solid state physics {such as} ferromagnetism, antiferromagnetism, the Mott transition, high-temperature superconductivity, Bose$-$Einstein condensate in cold optical lattice \cite{5',6,7, Kucab1}. 

{Despite being an approximation of the realistic electron interactions in a crystal, the results obtained with this model may explain insulating, magnetic, and even superconducting effects in solids, including 1D conductors}. 

{In this paper one-dimensional Hubbard model is being considered \cite{14}. Given the possbility of having exact solutions for this model \cite{12,13, Janek} its importance is additionally magnified by the possibility of further generalization to higher dimensions. We consider the Young orthogonal basis, using the Jucys-Murphy operators \cite{1,2,3, Okounkov, Lascoux} which represents the irreducible basis partially adapted to the total symmetry of the system.} These operators generate a maximal Abelian subalgebra in the group algebra $\mathbb C [\Sigma_N]$ of the symmetric group, {allowing us} to {interprete} the orthogonal basis of standard Young tableaux as that of common eigenstates of a complete set of commuting operators \cite{Cheng1, Cheng2}, in a {frame of Dirac's quantum-mechanical principles \cite{Dirac}.} 

{Considering that interacting particles are fermions, therefore, due to the exclusion principle, maximum of two particles are allowed to populate a single node.} The electrons moving along the ring with $N$ nodes labeled by the set $\tilde{N} =\{j=1,2,\dots ,N\}$ have two possibilities of the one-node spin projection, namely $+\frac{1}{2}$ (denoted by $+$) and $-\frac{1}{2}$ (denoted by $-$), given by the set $\tilde{2} =\{i=+,-\}$. {Since periodic boundary conditions are imposed} one has to assume that $j+N=j$. We introduce two {additional} sets $\tilde{N_{i}} =\{j_{i}=1,2,\dots ,N_{i}\}$, $i\in\{+,-\}$, with cardinality equal to the number of electrons with the spin projection in the set $\tilde{2}$, moving on the ring. The whole number of electrons is equal then $N_{+}+N_{-}=N_{e}$, and {is} labeled by the set $\tilde{N_{e}}=\{j_{e}=1,2,\dots , N_{e}\}$. These electrons can hop to the nearest neighbours, {provided the Pauli principle was not violated}, and interact when two are on the same node, thus {individual $N_i$ numbers are conserved}. Since we are dealing with the finite number of nodes of the lattice given by $N$, we assume the single band approximation.

\section{The Hubbard Hamiltonian and its symmetries}

{To get an insight into the} dynamics of finite system of interacting electrons occupying the one-dimensional chain consisting of $N$ atoms {we use the Hubbard Hamiltonian of the following form}
\be\label{1}
\hat{H}=t\sum_{i\in \tilde{2}}\sum_{j\in \tilde{N}}(\hat{a}^{\dagger}_{ji}\hat{a}_{j+1 i}+\hat{a}^{\dagger}_{j+1 i}\hat{a}_{ji})+U\sum_{j\in \tilde{N}}\hat{n}_{j \,\,+}\hat{n}_{j\,\,-},
\ee
where $\hat{n}_{ji}=\hat{a}^{\dagger}_{ji}\hat{a}_{ji}$, and $\hat{a}^{\dagger}_{ji}$ , $\hat{a}_{ji}$ are the canonical Fermi operators {of creation and anihilation of an electron of spin $i$, on the site $j$.} 

{The first component of the Hamiltonian $(\ref{1})$ is responsible for the \textit{wave-like} behaviour of the electrons, whereas the second component corresponds to the \textit{particle-like} behaviour provided electron-electron interactions occur and are described by a characteristic interaction constant denoted by $U$ \cite{Tasaki}.} In general $U$ can be {of} any value, with $U <0$ ($U<<0$ - the case presented in this article) and $U> 0$ ($U>>0$ \cite{16, Jak2}) are responsible for attraction and repulsion, respectively{.} $U=0$ stands for no effect {f}or plain gas of fermions.\\

The \emph{single-node} space $h_{j}$ has the basis consisting of $n$ vectors denoting all possible occupations of one node{.} {For the} fermions
\be\label{2}
\mbox {dim} \,h_{j}=n=4,\,\, h_{j}=lc_{\mathbb{C}}\{\pm, \emptyset , +, -\},
\ee
where $\emptyset$ denotes the empty node, $+$ and $-$ stand for one-node spin projection equal to $\frac{1}{2}$ and $-\frac{1}{2}$, respectively, $\pm$ denotes the double occupation of one node by two electrons with different spin projections, and $lc_{\mathbb{C}}A$ stands for the linear closure of a set $A$ over the complex field $\mathbb{C}$.

The final Hilbert space $\mathcal{H}$ of all quantum states of the system has the form
\be\label{3}
\mathcal{H}=\prod_{j=1}^{N} \otimes\,\, h_{j},\,\, \mathcal{H}=\sum_{N_{e}=0}^{2N}\oplus \,\, \mathcal{H}^{N_{e}},
\ee
where $\mathcal{H}^{N_{e}}$ denotes the space with fixed number of electrons $N_{e}$.

{The initial, orthonormal basis of the Hilbert space $\mathcal{H}$ consists of linearly independent vectors called \emph{electron configurations} \cite{Jakubczyk}, defined by the following mapping}
\be\label{4}
f:\tilde{N}\longrightarrow \tilde{4},\,\,\tilde{4}=\{\pm , \emptyset , +, -\},
\ee
and constitute the $N$-sequences of the elements from the set $\tilde{4}$
\be\label{5}
|f\rangle=|f(1)f(2)\dots f(N)\rangle=|i_{1} i_{2} \dots  i_{N}\rangle,\,\, i_{j}\in \tilde{4},\,\, j\in \tilde{N},
\ee
with
\be\label{6}
\tilde{4}^{\tilde N} = \{ f: \tilde N \longrightarrow \tilde 4 \},
\ee
\be\label{7}
\mathcal{H} = lc_{\mathbb{C}} \,\, \tilde 4^{\tilde N}.
\ee

{Previously, work on problems related to the symmetries of the one-dimensional Hubbard model has appeared in the literature} starting from Lieb and Wu \cite{8}, Yang \cite{9} and continued in, inter alia, Refs. \cite{Tasaki,Yang3,Yang4}, with the book of Essler et al. being the eminent summary and supplement of their work \cite{Essler}. 

Since the periodic boundary conditions are {set}, the Hamiltonian $(\ref{1})$ has the obvious translational symmetry $(\hat{a}_{N+1i}=\hat{a}_{1i})$, {therefore} one-particle Hamiltonian of the form $(\ref{1})$ is completely diagonalized by a Fourier transformation. Apart from the cyclic symmetry, system reveals{,} among others{,} two independent $SU(2)$ {symmetries \cite{Essler,Noga}, $SU(2)\times SU(2)$, in spin and pseudo-spin space \cite{Jak}.} This symmetry involves spin and charge degrees of freedom, thus one has two sets of generators, $\{\hat{S}_{z}, \hat{S}^{+}, \hat{S}^{-}\}$ and $\{\hat{J}_{z}, \hat{J}^{+}, \hat{J}^{-}\}$, for spin and charge respectively. These generators can be written in the following forms
\be\label{r6''}
\hat{S}_{z}=\frac{1}{2}\sum_{ j\in \tilde{N }}(\hat{a}^{\dagger}_{j+}\hat{a}_{j+}-\hat{a}^{\dagger}_{j-}\hat{a}_{j-}),\,\,\hat{S}_{+}={\hat{S}_{-}}^{\dagger}=\sum_{ j\in \tilde{N }}\hat{a}^{\dagger}_{j+}\hat{a}_{j-},
\ee
\be\label{r6'''}
\hat{J}_{z}=\frac{1}{2}\sum_{ j\in \tilde{N }}(\hat{a}^{\dagger}_{j+}\hat{a}_{j+}+\hat{a}^{\dagger}_{j-}\hat{a}_{j-}-1),\,\,\hat{J}_{+}=\sum_{ j\in \tilde{N }}(-1)^{j}\hat{a}^{\dagger}_{j+}\hat{a}^{\dagger}_{j-},\,\,\hat{J}_{-}=\sum_{ j\in \tilde{N }}(-1)^{j}\hat{a}_{j+}\hat{a}_{j-}
\ee
and the transfer between these two sets is known as the {Shiba transformation from the mapping rotations in spin space into rotations in pseudo-spin space, achieved by changing the sign of $U$ \cite{8,Essler}.} 

\section{Jucys-Murphy operators}

The operators $\hat{M}_{j}$, defined within the symmetric group algebra $\mathbb{C}[\Sigma_{N}]$ as the sum of all transpositions $(j,j')$ of the node $j\in \tilde{N}$ with preceding nodes $j'<j$, {were introduced by Jucys \cite{1,2} and independently by Murphy \cite{3}.} These {correspond to} $N-1$ hermitian and mutually commuting operators of the form
\be\label{9}
\hat{M}_{j}=\sum_{1\leq j' <j}(j,j'),\,\, j=2,3,\dots , N,
\ee
{which} generate a maximal Abelian subalgebra in $\mathbb{C}[\Sigma_{N}]$. The standard Young tableaux $|\lambda_{N} \, y>$ \cite{5} of the shape $\lambda_{N}\vdash N$, i.e. the tableaux of this shape in the alphabet $\tilde N$ of nodes, with strictly increasing entries in rows and columns, constitutes the common eigenvector $|\lambda_{N} \, y>$ of the set of $\hat{M}_{j}$ operators, that is
\be\label{9'}
\hat{M}_{j}|\lambda_{N} \, y>=m_{j}(y)|\lambda_{N} \, y>,
\ee
with eigenvalues
\be\label{9''}
m_{j}(y)=c_{j}(y)-r_{j}(y),
\ee 
where the pair of positive integers $(c_{j}(y),r_{j}(y))$ gives the positions (the column and the row) of the number $j$ in tableaux $|\lambda_{N} \, y>$. 

{This leads to} each basis function of the irreducible representation $\Delta^{\lambda_{N}}$ of the symmetric group $\Sigma_{N}$, labeled by the Young tableaux $|\lambda_{N} \, y>$, {being} completetly determined by the sequence $(m_{1}=1, m_{2},\dots , m_{N})$ of eigenvalues $(\ref{9''})$. The realization of each such irreducible vector within the group algebra $\mathbb{C}[\Sigma_{N}]$ is given via the {projection} operator $e_{yy}^{\lambda_{N}}$ of the well known Young orthogonal basis \cite{1,2,3}, where the {importance} of the Jucys-Murphy operators is underlined. 

\section{The irreducible basis}

Since the system of the half-filling magnetic ring reveals two independent $SU(2)$ symmetries \cite{Essler,Noga} { $SU(2)\times SU(2)$} - the spin and the charge degrees of freedom (called \emph{particles}) are possible in the spin and the pseudo-spin space, respectively. 

{Assuming there is only one particle on the ring we allow the symmetric group $\Sigma_{N}$ act on the number of nodes $N$}. The set of all allowed positions of {the} particle {at $N$ nodes} forms the orbit $O_{\mu}$ of the symmetric group $\Sigma_{N}$, labeled by the weight \cite{8} $\mu=(\mu_{1}, \mu_{2})$, where $\mu_{1}=N-1$ and{,} since there is only one particle in the system{,} $\mu_{2}=1$. Such an orbit is invariant under the action of the symmetric group $\Sigma_{N}$ and forms the carrier space of the transitive representation $R^{\Sigma_{N}:\Sigma^{\mu}}$, with the stabilizer $\Sigma^{\mu}$ being the Young subgroup $\Sigma^{\mu}=\Sigma_{\mu_{1}}\times \Sigma_{\mu_{2}}$, {where} $\times$ denotes the Cartesian product. 

One can {obtain states} with definite permutational symmetry of the $N_{p}$ particles of one kind {contained either within the spin space or within the pseudo-spin space} by taking the irreducible basis of the appropriate irreducible representations $\Delta^{\lambda_{N}}$ of the symmetric group $\Sigma_{N}$, where $\lambda_{N} \vdash  N$, in the tensor products of $N_{p}$ transitive representations $R^{\Sigma_{N}:\Sigma^{(N-1,1)}}$ along with the appropriate decomposition
\be\label{11}
R^{\Sigma_N : \Sigma^\mu} \cong \sum_{ \lambda_{N} \unrhd \mu} K_{\lambda_{N} \, \mu} \,\, \Delta^{\lambda_{N}}.
\ee
{Here} $K_{\lambda_{N} \, \mu}$ {corresponds to the} Kostka numbers \cite{20}, and the sum runs over all partitions $\lambda_{N}$ greater than or equal to $\mu$ in the dominance order \cite{Kerber}.

In order to obtain Young orthogonal basis \cite{1,2,3} of the tensor product of the appropriate transitive representations 
\be\label{12}
(R^{\Sigma_{N}:\Sigma^{(N-1,1)}})^{\otimes N_{p}},
\ee
with the decomposition into irreducible representations $\Delta ^{\lambda_{N}}$ given by equation $(\ref{11})$, we are going to use Jucys-Murphy operators and create  the projection operators
\be\label{12''}
e_{yy}^{\lambda_{N}} = | \lambda_{N} w y\rangle \langle \lambda_{N} w y| 
\ee
in the space of the tensor product $(h_{j})^{\lambda_{N_{p}}}${.} {Here} $w$ {is an} appropriate repetition label and $y \in SYT(\lambda_{N})$, with $SYT(\lambda_{N})$ being the set of all standard Young tableaux of the shape $\lambda_{N}$.

Thus we have
\be\label{p}
e_{yy}^{\lambda_{N}}=\prod_{j=2}^{N}\prod_{\{y_{j-1}|y_{j-1}^{+}\neq y_{j}\}}\frac{\hat{M}_{j}-m_{j}(y_{j-1}^{+})\hat{I}}{m_{j}(y)-m_{j}(y_{j-1}^{+})},
\ee
{with} $y \in SYT(\lambda_{N})$ {and} $y_{j}$ {corresponding to} the tableaux {obtained} from $y$ by extracting the set $\{j+1, j+2, \dots N\}$ of numbers{.} $y_{j-1}^{+}$ can be created from $y_{j-1}$ after adding to its entries the number $j$, and $\hat{I}$ stands for the appropriate unity operator.

\section{The example}

As {discussed} in the previous section the system of the half-filling magnetic ring reveals, among others, two independent $SU(2)$ symmetries, {allowing us} to use the {above technique} for the pseudo-spin space {\cite{Jak}}. 

The set of electron configurations for the attractive Hubbard model {corresponding to $U<<0$} \cite{Jakost}, {does} not contain the elements with two atoms singly occupied by opposite spin projection (unpaired spins) and thereby provides the charge degrees of freedom only. 

{Consider} the irreducible basis for the half-filling case of $N=4$, $N_{-}=1$, and $U<<0$ in the form
\be\label{14}
R^{\Sigma_{4}:\Sigma^{(3,1)}}\otimes R^{\Sigma_{4}:\Sigma^{(3,1)}}= R^{\Sigma_{4}:\Sigma^{(3,1)}}\oplus R^{\Sigma_{4}:\Sigma^{(2,1^2)}}
\ee
of the transitive representations with appropriate decomposition into irreducible representations $\Delta ^{\lambda}$ given {by}
\be\label{15}
R^{\Sigma_{4}:\Sigma^{(3,1)}}=\Delta^{\{4\}}\oplus \Delta^{\{3,1\}},
\ee
and
\be\label{16}
(\Delta^{\{4\}}\oplus \Delta^{\{3,1\}})\otimes (\Delta^{\{4\}}\oplus \Delta^{\{3,1\}})=2\Delta^{\{4\}}\oplus 3 \Delta^{\{3,1\}}\oplus \Delta^{\{2^2\}}\oplus \Delta^{\{2,1^2\}}.
\ee

{States} related to the first component on the right-hand side of the tensor product in decomposition $(\ref{14})$ describe the situation {corresponding to the occupancy of a single node by two particles} and {are therefore to be rejected due to impossibility of combining $\pm$ and $\emptyset$ together into the single one-node state}. 

The transitive representation $R^{\Sigma_{4}:\Sigma^{(2,1^2)}}$ contains the symmetric
\be\label{17}
(R^{\Sigma_{4}:\Sigma^{(2,1^2)}})_{st}=\Delta^{\{4\}}\oplus \Delta^{\{3,1\}} \oplus \Delta^{\{2^2\}},
\ee
and the antisymmetric
\be\label{18}
(R^{\Sigma_{4}:\Sigma^{(2,1^2)}})_{a}= \Delta^{\{3,1\}}\oplus \Delta^{\{2,1^2\}},
\ee
part of the tensor product $(\ref{14})$. The decompositions of selected elements of the irreducible basis of the symmetric group $\Sigma_{4}$ onto the electron configurations {are presented in} Table \ref{Tab}.
\begin{table}
\begin{center}
\begin{tabular}{|c|c|c|c|c|c|}
\hline
$|f\rangle$ & $\young(13,24)$ & $\young(12,34)$& $\young(12,3,4)$ & $\young(13,2,4)$ & $\young(14,2,3)$\\
\hline
$\pm \emptyset + +$ & $0$ & $\frac{1}{\sqrt{6}}$& $0$ & $\frac{1}{2\sqrt{3}}$ & $\frac{-1}{\sqrt{6}}$\\  
$\pm + \emptyset +$ & $\frac{1}{2\sqrt{2}}$ & $\frac{-1}{2\sqrt{6}}$& $\frac{1}{4}$ & $\frac{1}{4\sqrt{3}}$ & $\frac{1}{\sqrt{6}}$\\
$\pm + + \emptyset$ & $\frac{-1}{2\sqrt{2}}$ & $\frac{-1}{2\sqrt{6}}$& $\frac{-1}{4}$ & $\frac{-\sqrt{3}}{4}$ & $0$\\
$\emptyset \pm + +$ & $0$ & $\frac{1}{\sqrt{6}}$& $0$ & $\frac{-1}{2\sqrt{3}}$ & $\frac{1}{\sqrt{6}}$\\
$+ \pm \emptyset +$ & $\frac{-1}{2\sqrt{2}}$ & $\frac{-1}{2\sqrt{6}}$& $\frac{1}{4}$ & $\frac{-1}{4\sqrt{3}}$ & $\frac{-1}{\sqrt{6}}$\\
$+ \pm + \emptyset$ & $\frac{1}{2\sqrt{2}}$ & $\frac{-1}{2\sqrt{6}}$& $\frac{-1}{4}$ & $\frac{\sqrt{3}}{4}$ & $0$\\
$\emptyset + \pm +$ & $\frac{1}{2\sqrt{2}}$ & $\frac{-1}{2\sqrt{6}}$& $\frac{-1}{4}$ & $\frac{-1}{4\sqrt{3}}$ & $\frac{-1}{\sqrt{6}}$\\
$+ \emptyset \pm +$ & $\frac{-1}{2\sqrt{2}}$ & $\frac{-1}{2\sqrt{6}}$& $\frac{-1}{4}$ & $\frac{1}{4\sqrt{3}}$ & $\frac{1}{\sqrt{6}}$\\
$+ + \pm \emptyset$ & $0$ & $\frac{1}{\sqrt{6}}$& $\frac{1}{2}$ & $0$ & $0$\\
$\emptyset + + \pm$ & $\frac{-1}{2\sqrt{2}}$ & $\frac{-1}{2\sqrt{6}}$& $\frac{1}{4}$ & $\frac{\sqrt{3}}{4}$ & $0$\\
$+ \emptyset + \pm$ & $\frac{1}{2\sqrt{2}}$ & $\frac{-1}{2\sqrt{6}}$& $\frac{1}{4}$ & $\frac{-\sqrt{3}}{4}$ & $0$\\
$+ + \emptyset \pm$ & $0$ & $\frac{1}{\sqrt{6}}$& $\frac{-1}{2}$ & $0$ & $0$\\
\hline
\end{tabular}
\caption{The irreducible basis (columns 2 - 6) of the representations $\Delta^{\{2^2\}}$ and $\Delta^{\{2,1^2\}}$ of the symmetric group $\Sigma_{4}$ taken as the linear combinations of the electron configurations of the first column.}\label{Tab}
\end{center}
\end{table}

{According to (\ref{17}) and (\ref{18}) application of the irreducible basis to the $12$-dimensional Hubbard Hamiltonian for the case of $N=4=N_{e}$, $N_{-}=1$, leads to the quasidiagonal form with three three-dimensional, and three one-dimensional blocks on the diagonal.}

\section{Conclusions}
We have demonstrated the applications of the Young irreducible basis in the diagonalization procedure of the one-dimensional Hubbard {model}, using the set of Jucys-Murphy operators. 

We have shown the way of finding the basis with definite permutational symmetry based on the properties of the maximal Abelian subalgebra in the group algebra $\mathbb C [\Sigma_N]$ of the symmetric group $\Sigma_{N}$. 

We {explored the example of an attractive Hubbard model} ($U<<0$) and the half-filling magnetic rings with $N$ nodes occupied by $N_{e}=N$ electrons, including $N-1$ electrons with the same spin projection. 

This approach {leads to a significant reduction in the size of the Hubbard Hamiltonian and the result obtained} for the projection operators of the Young orthogonal bases {can easily be implemented into numerical simulations due to the use of  simple transpositions being the generators of the symmetric group.}

\end{document}